\documentclass{amsart}
\usepackage{amssymb}

\usepackage{graphicx,epstopdf}

\newtheorem{theorem}{Theorem}
\newtheorem{lemma}{Lemma}
\newtheorem{prop}{Proposition}
\newtheorem{corollary}{Corollary}
\newtheorem{conj}{Conjecture}

\begin{document}
\title
{Elliptic Faulhaber polynomials and Lam\'{e} densities of states}%
%
\author{M-P. Grosset}
\address{Department of
Mathematical Sciences,
          Loughborough University,
Loughborough,
          Leicestershire, LE11 3TU, UK}
          \email{M.Grosset@lboro.ac.uk}
          
         \author{A.P. Veselov}
\address{Department of
Mathematical Sciences,
          Loughborough University,
Loughborough,
          Leicestershire, LE11 3TU, UK
          and
Landau Institute for Theoretical Physics, Moscow, Russia}

\email{A.P.Veselov@lboro.ac.uk}

\maketitle
\begin{abstract}
A generalisation of the Faulhaber polynomials and Bernoulli numbers related to elliptic curves is introduced and investigated. This is  applied to compute the density of states for the classical Lam\'e operators.
\end{abstract}


\section{Introduction}
Johann Faulhaber, "Rechenmeister der Stadt Ulm", in his book \cite{Faul} published in 1631 announced a remarkable fact that the sums of the odd powers of the first $n$ natural numbers can be expressed as the polynomials of their sum. Namely, if we denote
$S_1 = 1 + 2 + ... + n = (n^2 + n)/2$ as $\lambda$ then
$$S_3 = 1^3 + 2^3 + ... + n^3  = \lambda^2,$$
$$S_5 = 1^5 + 2^5 + ... + n^5  = (4\lambda^3 - \lambda^2)/3,\\$$
$$S_7 = 1^7 + 2^7 + ... + n^7  = (12\lambda^4 - 8\lambda^3 + 2\lambda^2)/6 $$
and in general
$$S_{2m-1} = 1^{2m-1} + 2^{2m-1} + ... + n^{2m-1} = F_m(\lambda)$$
for some polynomials $F_m,$ which are called {\it Faulhaber polynomials}:
$$F_1(\lambda) = \lambda, \quad F_2(\lambda) = \lambda^2, \quad F_3(\lambda) = \frac{1}{3} \lambda^2(4\lambda-1),\quad F_4(\lambda) = \frac{1}{3} \lambda^2(6\lambda^2-4\lambda +1),$$
$$F_5(\lambda) = \frac{1}{5} \lambda^2(16\lambda^3-20\lambda^2 +12\lambda -3),\quad
F_6(\lambda) = \frac{1}{3} \lambda^2(16\lambda^4-32\lambda^2 +34\lambda^2 -20\lambda+5).$$  In full generality this claim was first proved by  Jacobi in 1834 \cite{J}. For a nice discussion of this story
and the effective ways to compute the Faulhaber polynomials we refer
to a very interesting paper \cite{K} by Donald Knuth.

For the general $k$ the sums of powers $S_{k-1} (n) = 1^{k-1} + 2^{k-1} +\dots +n^{k-1}$  can be expressed through the classical {\it Bernoulli polynomials} (see e.g. \cite{Er}):
$$S_{k-1}(n) = (B_{k}(n+1)-B_{k})/k,$$
where $B_k = B_k(0)$ are the {\it Bernoulli
numbers}. Bernoulli polynomials can be defined through the generating function:
$$\frac{ze^{zx}}{e^z-1}=\sum_{k=0}^{\infty}\frac{  B_{k}(x)}{k!}z^k$$
and have the symmetry property
$B_k(1-x) = (-1)^k B_k(x).$  Faulhaber's claim is a simple corollary of this symmetry.
In fact, Faulhaber polynomials are related to the even Bernoulli polynomials in a simple way:
$$
B_{2m}(x+1) = 2m F_m\left(\frac{x^2+x}{2}\right) + B_{2m}.
$$
Faulhaber did not discover Bernoulli numbers but his work was known to Jacob Bernoulli and cited in his famous treatise in {\it Ars Conjectandi} in 1713.

Recently, D. Fairlie and one of the authors \cite{FV} discovered an interesting relation of the Faulhaber and Bernoulli polynomials with the theory of
the Korteweg-de Vries (KdV) equation
$$u_t - 6uu_x + u_{xxx} = 0.$$
It is known since 1967 due to Gardner, Kruskal, Miura and Zabusky \cite{Miura II, Miura V} that this equation has infinitely many conservation laws of the form
\begin{equation}
\label{Kdvint}
I_m[u] = \int T_m (u, u_x, u_{xx}, ..., u_{m-2}) dx, 
\end{equation}
where $T_m[u]$ are some polynomials of the function $u$ and its $x$-derivatives up to order $m-2:$
$$ T_1  =   u, \quad  T_2 =   u^2, \quad T_3 =   {u_1}^2 + 2 u^3, \quad T_4= {u_2}^2 + 10 u {u_1}^2 + 5 u^4...$$
They are uniquely defined by some homogeneity property modulo adding a total derivative
and multiplication by a constant (see the next section).
This constant can be fixed by demanding that $T_m (u, u_x, u_{xx}, ..., u_{m-2})  = u_{m-2}^2$ plus a function of derivatives of order less than $m-2.$

The KdV equation has famous solutions known as {\it solitons}, the simplest of which is a one-soliton solution $$ u = - 2 {\rm \mbox{sech}}^2 (x - 4t), $$
corresponding to the initial profile $u(x,0) = -2 {\rm \mbox{sech}}^2 x. $
The main result of \cite{FV}  is the following formula, relating the Faulhaber polynomials $F_m$
with the integrals of the KdV equation:
$$
I_{m}[-2 \lambda \mbox{sech} ^2 x]= (-1)^{m}\frac{2^{2m}}{2m-1} F_{m}(\lambda).
$$

In the present paper we introduce a new class of polynomials, which we call
{\it elliptic Faulhaber polynomials.} They are related to the periodic traveling waves (known also as {\it cnoidal waves}) of the KdV equation:
$$ u(x,t) =  2 \wp_*(x-ct),$$ where $\wp_*(x)$ satisfies the differential equation
\begin{equation}
(\wp_*') ^2
 = 4\wp_* ^3 - g_1\wp_*  ^2 - g_2 \wp_* -g_3
\label {generalised d.e}
\end{equation}
with $g_1 = - c.$
The function $\wp_*(x)$ differs from the classical Weierstrass elliptic $\wp$-function by adding a constant.

The elliptic Faulhaber polynomials are defined as
\begin{equation}
\label{elliptic Faulhaber}
\mathcal{F}_m(\lambda | \Gamma, \gamma) = \oint _\gamma T_{m}[2
\lambda \wp_*(z)] dz,
\end{equation} 
where the integral here is taken over a cycle
$\gamma$ on the corresponding
elliptic curve $\Gamma$ given by the equation
$Y^2 = 4 X^3 - g_1 X^2 - g_2 X - g_3$ with $X= \wp_*(z), Y = \wp_*'(z).$
Note that since the integrand has all the residues zero, one can consider $\gamma$ here as an element
of the first homology group $H_1 (\Gamma, \bf{Z})$ and the integral is a linear function on it.
Equivalently, one can say that the integrand is a differential of a second kind and thus, by the de Rham theorem, determines a special element of the first cohomology group $H^1 (\Gamma, \bf{C}),$ polynomially dependent on the parameter $\lambda.$

 Let $$\omega = \frac{1}{2} \oint_\gamma dz =  \frac{1}{2} \oint_\gamma \frac{dX}{Y} , \quad \xi = -\frac{1}{2} \oint_\gamma \wp^*(z) dz= -\frac{1}{2} \oint_\gamma  \frac{X dX}{Y}$$
be the corresponding periods of the basic second-kind differentials on this curve. The variables $g_1, g_2, g_3, \omega, \xi$ can be considered as the coordinates on the space $\mathcal{U}= \{(\Gamma, \gamma)\},$ so $\mathcal{F}_m= \mathcal{F}_m(\lambda; g_1, g_2, g_3, \omega, \xi) .$ Define the {\it weight} of the variables
$\lambda, g_1, g_2, g_3, \omega,\xi$ as $0, 2, 4, 6, -1, +1$ respectively. 

\begin{theorem} 
The elliptic Faulhaber polynomial  $\mathcal{F}_m$ is polynomial in all variables $\lambda, g_1,g_2, g_3, \omega, \xi$
with rational coefficients, homogeneous with weight $2m+1.$
When $g_2=g_3=0$ it reduces (up to a factor) to the classical Faulhaber polynomial
\begin{equation}
\label{Fsol}
\mathcal{F}_m( \lambda; g_1, 0, 0, \omega, \xi) =-\frac{4}{2m-1} g_1^{m-1}\xi F_m (\lambda) .
\end{equation}
\end{theorem}

Here are the first 4 elliptic Faulhaber polynomials:
$$\mathcal{F}_1 = - 4 \xi  \lambda,  \quad \mathcal{F}_2 = (-\frac{4}{3} g_1 \xi + \frac{2}{3} g_2 \omega) \lambda^2,$$
$$ \mathcal{F}_3 = ( -\frac{4}{15} g_1^2 \xi + \frac{2}{15} g_1g_2 \omega) \lambda ^2 ( 4 \lambda -1)- \frac{8}{5} g_2 \xi \lambda ^2 ( 3 \lambda -2) + \frac{8}{5} g_3  \omega  \lambda ^2 (2\lambda -3),$$   
$$\mathcal{F}_4 = ( \frac{4}{21}  g_1^3  \xi - \frac{2}{21} g_1^2 g_2 \omega) \, \lambda ^2 ( 6 \lambda ^2 - 4 \lambda + 1) - \frac{8}{21} g_1g_2  \xi  \, \lambda ^2 ( 26 \lambda ^2 - 29 \lambda + 9)$$
$$+ \frac{8}{7}  g_1g_3  \omega \lambda ^2 ( 3 \lambda ^2 - 2 \lambda -3)+ \frac{2}{21}  g_2^2  \omega \lambda ^2 ( 25 \lambda ^2 - 40 \lambda + 24)] 
- \frac{32}{7}  g_3 \xi \lambda ^2 (5 \lambda ^2 -15\lambda +9).
$$

If we consider the Weierstrass reduction $g_1=0$ we come to the reduced version
of these polynomials $F_m^W(\lambda; g_2, g_3, \omega, \eta) = \mathcal{F}_m( \lambda; 0, g_2, g_3, \omega, \eta),$ where $\wp^* = \wp,$  $\xi$ reduces to $\eta = -\frac{1}{2} \oint \wp(z) dz = \zeta(\omega; g_2, g_3),$ $\wp$ and $\zeta$ are the 
standard Weierstrass elliptic functions \cite{Whittaker}. One can see the explicit form of the first 8 of them in the Appendix.

These polynomials have double zero at zero (for $m>1$) with the second derivative at zero proportional to 
$$
\mathcal{B}_{2m} = \frac{1}{2} \oint_\gamma (\frac{d^{m-1}}{dz^{m-1}} \wp(z) )^2 dz, 
$$
which we call {\it elliptic Bernoulli numbers.} Note that in contrast to the so-called Bernoulli-Hurwitz numbers $BH_{2k}$  \cite{Hurwitz, Katz}, which are functions of only $g_2, g_3,$ these numbers $\mathcal{B}_{2m}= \mathcal{B}_{2m}(g_2, g_3, \omega, \eta)$ depend also on $\omega$ and $\eta$: $$\mathcal{B}_{2} =   3 g_2 \omega, \quad
\mathcal{B}_{4} =  - \frac{3}{5} g_3 \omega + \frac{2}{5} g_2 \eta, \quad  \mathcal{B}_{6} =  \frac{2}{7} g_2^2 \omega - \frac{36}{7} g_3 \eta.$$
On the discriminant (when two of the roots of the equation $4 X^3 - g_2 X - g_3=0$ collide) they reduce up to a simple factor to the usual Bernoulli numbers:
$$\mathcal{B}_{2n} (\frac{1}{12} g_1^2,\, \frac{1}{216} g_1^3, \,\omega, \, \xi + \frac{1}{12} g_1 \omega) = - B_{2n} \, g_1^n \xi.$$


Our main motivation for considering these polynomials came from the theory of the classical {\it Lam\'e equation}
$$ - y'' + n(n+1) \wp(z) y = E y,$$
well known since XIX century (see \cite{Er, Whittaker}). The classical procedure going back to Hermite and Halphen allows one to find the general solution $y(z)$ for given $n=1, 2, 3....$ (see \cite{BE, Take1} for the recent results in this direction). Our approach is somewhat different: we want to say something about
the Lam\'e equation for {\bf all} values of the parameter $n.$

More precisely, we consider the {\it density of states} $\rho(E)$ for the corresponding Lam\'e operator 
$L= - \frac{d^2}{dx^2} +
n(n+1)\wp(x).$ The density of states is one of the most important spectral characteristics of the Schr\"odinger operator with periodic potential (see e.g.  \cite{Shenk}). From the finite-gap theory \cite{Ince, Nov} it is known that the density of states of the Lam\'e operator with integer $n$  has the form
\begin{equation}
\label{ro}
\rho(E) =\frac{1}{2 \pi }  \frac{P_n(E)}{ \sqrt{R_{2n+1}(E)}}, 
\end{equation}
for certain polynomials $P_n, R_{2n+1}$. 
On the other hand, the  high-energy asymptotics for $\rho$ can be given in terms of the KdV integrals (see \cite{Nov}), which in this case are our (reduced) elliptic Faulhaber polynomials:
$$
\rho(E) = \frac{1}{2 \pi \sqrt E} [1+ \frac{1}{T} \sum_{k=1}^{\infty} \frac{ 2k-1}{2^{2k-1}}\frac{ F_{k}^W(\lambda)}{ E^{k}}], \quad \lambda = \frac{n(n+1)}{2}.
$$
Let $\bar{\wp}=-\frac{\eta}{\omega}$ be the average of the Weierstrass $\wp$-function over real period $T=2\omega$. We have the following

\begin{theorem} The coefficient $a_k = a_k(n)$ in the numerator $P_n(E) = E^n + a_1 E^{n-1} + a_2 E^{n-2} + \dots +a_n$ of the Lam\'e density of states (\ref{ro}) is a polynomial in $n$ of degree $[\frac{5k}{2}].$ Its coefficients are polynomials with rational coefficients of $g_2, g_3$ and $\bar{\wp},$ homogeneous of weight $2k,$ where the weights of $\bar{\wp}, g_2, g_3$ are 2, 4, 6 respectively.
\end{theorem}

The procedure is actually effective and gives the explicit form of the first coefficients $a_k(n)$ for all $n$:
\[a_1=\frac{n (n+1)}{2} \bar{\wp},\]
\[ a_2 = -\frac{g_2}{480}(n-1) n(n+1)(6+25 n +16 n^2),\]
\[a_3= (n-2)(n-1)n(n+1)(-\frac{g_3}{3360} (45 + 243n + 247 n^2+ 64 n^3) - \frac{g_2\bar{\wp}}{960}n(n+1)(27 + 16 n))\]
More of these formulas, which we believe to be new, are given in section 6 of this paper.

In the last section we discuss also some open problems and conjectures about the elliptic Faulhaber polynomials. 

\section{The densities of KdV integrals \label{KdV densities}}

The KdV integrals $I_k$ play a fundamental role in the theory of the 
one-dimensional Schr\"odinger operator $L= - \frac{d^2}{dx^2} +
u(x).$ They appear in the asymptotic expansion of the trace of the
resolvent of $R = (\mu I - L)^{-1}$ (see Gelfand-Dikii \cite{GD})
or, in the periodic case, of the density of states or quasi-momentum 
(see Novikov et al \cite{Nov} and section 6
below). In this section we will discuss the algebraic properties
of the corresponding densities $T_k$ mainly following the
classical paper by Miura, Gardner and Kruskal \cite{Miura V}. We
should warn the reader that our $u$ differs from the variable
in that paper by a constant factor, so the arithmetic of the
coefficients is slightly different.

The densities of the KdV integrals $T_k = T_k(u_0, u_1, ...,
u_{k-2})$ are polynomials in $u=u_0$ and its derivatives $u_x=u_1,
u_{xx} = u_2, ...$ up to order $k-2.$ In general they are defined
up to a total derivative, which does not affect the integrals. To
eliminate this freedom, let us introduce the notion of canonical
form \cite{Miura V}. Consider the following grading: for the monomial $u_0^{m_0}u_1^{m_1}...u_k^{m_k}$ we define its {\it rank} as $$ r=\sum_{j=0}^{j=k} (2 + j)m_j.$$
In other words, the rank of $u$ is 2 and each derivative gives additional $+1$.

A form $T(u_0, u_1, ..., u_{r})$ is called {\it
canonical} if it is rank-homogeneous and is irreducible in the
sense that there are no monomials
$u_0^{i_0}u_1^{i_1}...u_k^{i_k}$  in $T$ with the highest derivative $u_k$
appearing linearly (i.e. $i_k=1$).

It is clear that if the  highest derivative $u_k$ appears linearly
one can always eliminate such a term by doing integration by parts
(which is equivalent to adding a total derivative).

\begin{theorem} (\it{Miura, Gardner and Kruskal \cite{Miura V}}) For any $k=1,2,....$ the KdV equation has a non-trivial integral $I_k$ with a density
$T_k = T_k(u_0, u_1, ..., u_{k-2}),$ which is a rank-homogeneous polynomial of rank $2k$.
The canonical form of the corresponding density $T_k$
is defined uniquely up to a constant factor.
\end{theorem}

The highest derivative term $u_{k-2}$ of the corresponding $T_k$
is $u_{k-2}^2.$ We normalise these densities by choosing the
coefficient at this term to be 1. It turns out that this makes
all the other coefficients integer, which will be important for us.

\begin{theorem} All the coefficients of the KdV density $T_k$
in the normalised canonical form are integers.
\end{theorem}

Although this fact was probably known to the experts in KdV theory,
it did not play any role before. We have not found the proof in the
literature although it is not difficult.

{\it Proof of theorem 4.}  One can define the KdV densities by
the following recurrent formula:
\begin{equation}
\sigma_{m+1}=-\sigma_{m}^{'} - \sum_{k=1}^{k=m-1} \sigma_k
\sigma_{m-k} \label{density recurrence}
\end{equation}
with $\sigma_1 = u$ (see e.g. \cite{Nov}). These densities are not
in the canonical form but are rank-homogeneous and obviously have
integer coefficients. The even $\sigma$ are known to be total derivatives \cite{Nov}, so only odd $\sigma$ lead to non-trivial integrals. The claim is that
\begin{equation}
T_k = (-1)^{k-1} \sigma _{2k-1}^* ,\label{conserved density}
\end{equation}
where the polynomial $\sigma_{2k-1}^*$ is the irreducible
equivalent of  $\sigma_{2k-1}.$ We should only show that the
coefficient at $u_{k-2}^2$ in $T_k$ is equal to 1.

 From (\ref{density recurrence}) we have $\sigma_{2k-1}=-\sigma_{2k-2}^{'} - \sum_{i=1}^{i=2k-3} \sigma_i \sigma_{2k-2-i}.$
 The total derivative term $\sigma_{2k-2}^{'} $ vanishes when we take the irreducible form $\sigma_{2k-1}^{*}$ and so, does not
 contribute to the term  $u_{k-2}^2$. The only contributions to $u_{k-2}^2$ come therefore from $-\sum_{i=1}^{i=2k-3} \sigma_i \sigma_{2k-2-i}.$
Since the only  term of degree 1 in $\sigma_i$  is $(-1)^{i-1}u_{i-1}$
the contribution to $ u_{k-2}^2$ comes from $-\sum
_{i=1}^{i=2k-3}(-1)^{i-1}u_{i-1}(-1)^{2k-3-i}u_{2k-3-i}= -u_{k-2}^2 -2 \sum _{i=0}^{i=k-3}
u_iu_{2k-4-i}.$ After integration by parts $ u_i u_{2k-4-i}$ is reduced
to $(-1)^{k-i-2}u_{k-2}^2,$ so the quadratic term $u_{k-2}^2$ in
$\sigma_{2k-1}^*$ has the coefficient $-1-2(-1)^k\sum_{i=0}^{i=k-3}(-1)^i =(-1)^{k-1}$ for all $k.$ The theorem is proved.

Below are the first KdV densities (in the canonical form):
 \[ T_1  =   u\]
\[ T_2 =   u^2\]
\[T_3 =   {u_1}^2 + 2 u^3\]
\[ T_4= {u_2}^2 + 10 u {u_1}^2 + 5 u^4\]
\[ T_5  =  {u_3}^2+14u{u_2}^2 +70{u}^2{u_1}^2 +14{u}^5\]
\[ T_6 =   {u_4}^2 -20{u_2}^3+18 u{u_3}^2-35 {u_1}^4+126 {u}^2{u_2}^2 +420{u}^3{u_1}^2+42  {u}^6.\]

The following proposition gives an explicit formula for the last
coefficient in these polynomials.

\begin{prop}
The coefficient of $u_0^r$ in the canonical form $T_k$ is $\frac{2(2k-3)!}{k!(k-2)!}$
for $k>1$ and 1 for $k=1$.
\end{prop}

Proof. The term $u_0^k$ appears with exactly the same coefficient (up to a sign) 
in $T_k$ as in $\sigma_{2k-1}$ since the
integration by parts does not affect it. Consider again the
relation $\sigma_{2k-1}=-\sigma_{2k-2}^{'} - \sum_{i=1}^{i=2k-3}
\sigma_i \sigma_{2k-2-i}.$ The total derivative term
$\sigma_{2k-2}^{'} $  does not contain any term $u_0^k$. The only
contributions to $u_{0}^k$ come therefore from
$-\sum_{i=1}^{i=2k-3} \sigma_i \sigma_{2k-2-i}$. This leads to the following recurrence relation for the coefficients $b_k$ in $ T_k
= b_k u_0^r +.....:$ $$b_k= \sum_{i=1}^{i=k-1} b_i b_{k-i}$$
with $b_1 =1.$ For the generating function $G(t) =
\sum_{k=1}^{k=\infty} b_k t^k$ this means that $G(t)^2 - G(t) +t =0,$ i.e. $G(t) = \frac
{1 - \sqrt{1-4t}}{2}.$  Expanding $G(t)$ as a power series and equating the coefficients, we obtain
the required expression for $b_k.$

\section{The elliptic Faulhaber polynomials \label{EFP}}

Consider now the special potential of the form $u = 2 \lambda \wp_*,$ where
 $\wp_* = \wp_*(z;g_1,g_2,g_3)$ is the elliptic function satisfying the differential equation
$$(\wp_*') ^2
 = 4\wp_* ^3 - g_1\wp_*  ^2 - g_2 \wp_* - g_3.$$
This function differs from the classical Weierstrass function $\wp(z; \hat g_2, \hat g_3)$ with the equation
$$(\wp')^2= 4\wp^3-\hat g_2\wp - \hat g_3,$$
where $\hat g_2 = g_2 + \frac{1}{12} g_1 ^2$ and $\hat g_3 =  g_3 +\frac{1}{12} g_1 g_2 + \frac{1}{216}g_1^3,$ by adding a constant:
$$\wp_*(z; g_1,g_2,g_3)= \wp(z; \hat g_2, \hat g_3) + \frac{g_1}{12}.$$
However it will be convenient for us to keep the additional parameter $g_1$ 
and consider the elliptic curve $\Gamma$ in the non-reduced form
$$Y^2 = 4X^3 - g_1 X^2 -g_2 X - g_3.$$

Choose any cycle $\gamma$ on $\Gamma,$ which does not pass through the poles of the function $\wp_*$ and define the elliptic Faulhaber polynomials as the integrals
\begin{equation}
\label{elliptic Faulhaber}
\mathcal{F}_m(\lambda | \Gamma, \gamma) = \oint_\gamma T_{m}[2
\lambda \wp^*(z)] dz.
\end{equation} 
Since the integrand has all the residues zero, this integral can be considered as a linear function on the first homology group of $\Gamma$ or equivalently as an element of the first cohomology group $H^1(\Gamma, \bf{C})$. Thus, the elliptic Faulhaber polynomials can be considered as $\lambda$-dependent sections of the canonical cohomology bundle over the space of the elliptic curves $\mathcal{E} = \{\Gamma\}.$

We choose the following basis in the first cohomology of $\Gamma:$
\begin{equation}
\label{basis}
\omega = \frac{1}{2} \oint dz =  \frac{1}{2} \oint \frac{dX}{Y}, \quad \xi = -\frac{1}{2} \oint \wp^*(z) dz = -\frac{1}{2} \oint \frac{XdX}{Y}.
\end{equation}

Now we are going to prove our Theorem 1 claiming that $\mathcal{F}_m = \mathcal{F}_m(\lambda, g_1,g_2, g_3, \omega, \xi)$ are polynomial in all variables with rational coefficients. 

{\it Proof of Theorem 1.}  We will need the following lemma about the derivatives of the function $\wp_*,$
which can be easily proved by induction using the differential equation (\ref{generalised d.e}) for the function $\wp_*$ and its corollary
\begin{equation}
\label{p_*''}
\wp_*'' = 6 \wp_*^2 - g_1 \wp_* -\frac{1}{2} g_2.
\end{equation}

\begin{lemma} 
The derivatives of the  elliptic function $\wp_*$ have the form
\begin{equation}
\label{wp*}
\wp_*^{(2k)} = A_k^* (\wp_*; g_1,g_2, g_3), \quad \wp_*^{(2k+1)} = {A^*_k}' (\wp_*; g_1,g_2, g_3) \wp_*'
\end{equation}
for some polynomials $A_k^*$ with rational coefficients.
The polynomials $A_k^*$ satisfy the following recurrence relation
\begin{equation}
\label{recur}
A_{k+1}^* = (4\wp_*^3-g_1 \wp_*^2-g_2\wp_* - g_3) {A^*_k}'' + ( 6 \wp_*^2 -g_1\wp_*- \frac{1}{2} g_2) {A^*_k}',
\end{equation}
where the derivative of $A^*_k$ is taken with respect to  $\wp_*.$
\label{p_*^{(n)}}
\end{lemma}

We know that the KdV density $T_{m}(u_0,...,u_{m-2})$ is a sum of monomials
$u_0^{m_0}u_1^{m_1}...u_k^{m_k}$ with rank $$ r=\sum_{j=0}^{j=k} (2 + j)m_j = 2m.$$ From lemma 1 it follows that if we substitute in such a monomial $u = 2\lambda \wp_*(x)$ we will have 
an expression of the form $Q(\wp_*,g_1, g_2, g_3) \lambda^{M}$, where $M= \sum_{j=0}^{j=k} m_j$ and $Q$ is a polynomial in $\wp_*, g_1,g_2, g_3$ with rational coefficients. Indeed, the fact that the rank of $T_k$ is even means that the first derivative of $\wp_*$ appears only in even powers and thus can be eliminated using the equation (\ref{generalised d.e}). 
It is easy to see also that if we define the grading of $\wp_*,g_1, g_2, g_3$ to be 2,2,4 and 6 respectively then $Q$ is homogeneous of degree $2m.$ 

Let us consider now the integrals $$K_n^*= \oint_\gamma \wp_* ^n (z)dz.$$ 

\begin{prop}
The integrals $K_n^*$ satisfy the following recurrence relation of the third order
\[ (8n-4)K_n^*= (2n-2) g_1K_{n-1}^*+ (2n-3) g_2 K_{n-2}^* + (2n-4) g_3 K_{n-3}^*\]
with initial terms $K_0^*= 2\omega, K_1^*= -2 \xi, K_2^*=\frac{1}{6}g_2 \omega - \frac{1}{3} g_1 \xi.$
\end{prop}

Indeed, using (\ref{p_*''}) we have $K_n^*=\oint \wp_*^{n-2}(z)\wp_*^2(z) dz= \oint \wp_*^{n-2}(z)(\frac{1}{6}\wp_*''(z)  +\frac{g_1}{6}  \wp_* +\frac{g_2}{12})dz= \frac{g_1}{6} \oint \wp_*^{n-1}(z) dz+\frac{g_2}{12}\oint \wp_*^{n-2}(z) dz+ \frac{1}{6} \oint \wp_* ^{n-2}(z)\wp_*'' (z) dz.$
Integrating by parts the third integral gives $\oint \wp_* ^{n-2}(z)\wp_*'' (z) dz =-(n-2) \oint \wp_* ^{n-3} (z) (\wp_*' (z))^2dz =-(n-2)\oint \wp_*^{n-3} (z)(4\wp_*^3 (z) -g_1 \wp_*^2(z)-g_2\wp_* (z) -g_3)dz=-(n-2)(4 K_n^* -g_1 K_{n-1}^*-g_2 K_{n-2}^* - g_3 K_{n-3}^*).$ Therefore $K_n^* = \frac{g_1}{6}K_{n-1}^* +\frac{g_2}{12}K_{n-2}^* - \frac{n-2}{6}(4 K_n^* -g_1 K_{n-1}^*-g_2 K_{n-2}^* - g_3 K_{n-3}^*)  =  - \frac{2n-4}{3} K_n^*+\frac{n-1}{6}g_1 K_{n-1}^* +\frac{2n-3}{12}g_2K_{n-2}^* +\frac{n-2}{6}g_3 K_{n-3}^*,$ which leads to the required recurrence relation. The expression for $K_0^*, K_1^*$ are obvious;
the form of $K_2^*$ follows from the relation (\ref{p_*''}).

\begin{corollary} The integrals $K_n^*$ have the form
\begin{equation}
\label{km*}
K_m^* = A_*^{(m)} (g_1,g_2, g_3) \omega - B_*^{(m)} (g_1,g_2, g_3) \xi,
\end{equation}
where $A_*^{(m)} (g_1,g_2, g_3),  B_*^{(m)} (g_1,g_2, g_3)$ are polynomials with positive rational coefficients.
\end{corollary} 

Here are the next few integrals:
\[ K_3^*=\frac{1}{30}(g_1g_2+6g_3) \omega -\frac{1}{30}(2g_1^2+9g_2)\xi,\]
\[ K_4^*=\frac{1}{840}(6g_1^2g_2 +25g_2^2+36g_1g_3) \omega -\frac{1}{210}(3g_1^2+26g_1g_2+60g_3)\xi,\]
\[ K_5^*=\frac{1}{2520}(4g_1^3g_2+33g_1g_2^2+24g_1^2g_3+168g_2g_3) \omega -\frac{1}{2520}(8g_1^2+102g_1^2g_2+147g_2^2+300g_1g_3)\xi.\]

Combining all this we have the proof of the first part of the theorem.
To prove the second part let us consider the case $g_2=g_3=0.$ The equation (\ref{generalised d.e}) then becomes simply
\[(\wp_*') ^2
 = 4\wp_* ^3 - g_1\wp_*  ^2\]
with the solution
$\wp_*(z)=\frac{\alpha^2}{\sinh^2(\alpha z)},$
where $\alpha$ is related to $g_1$ by
$g_1 = - 4 \alpha^2.$

For the vanishing cycle the integral $\xi= -\frac{1}{2}\oint \wp_*(z)dz$ is identically zero while  the integral  $\xi = -\frac{1}{2} \int_{-\infty}^{+\infty} \wp^*(z) dz$ taken over the real line and  corresponding to the diverging period $\omega$  has the finite value $\alpha.$
In fact, it is convenient to shift the real line in the complex plane by $i \frac {\pi}{2},$
then the function becomes $\wp_*(x + i \frac {\pi}{2}) = - \alpha^2 \mbox{sech}^2 \alpha x,$ which (up to a coefficient $\frac{1}{2}$) is the soliton profile (see the Introduction).

Now we can apply the result by Fairlie and one of the authors \cite{FV}, which claims that
\begin{equation}
I_{m}[-2 \lambda \mbox{sech} ^2 x]= (-1)^{m}\frac{2^{2m}}{2m-1} F_{m}(\lambda).
\label{KdV_Faulhaber}
\end{equation}
Let us remind here the idea of the proof. Consider the solution of the KdV equation
$$u_t - 6uu_x + u_{xxx} = 0$$ with initial data
$u(x,0) = -n(n+1) \mbox{sech}^2 x.$ It is well-known that for integer $n$ 
$$u(x,t) \sim -2 \sum_{k=1}^{n} k^2 \mbox{sech}^2 k(x-4k^{2}t-x_k)$$
as $t \to \infty$ for some constants $x_k$ (see e.g. \cite{DJ}, page 79).
Now we can use the fact that the integrals $I_m$ are actually the conserved quantities, so $I_m[u(x,t)] = I_m[u(x,0)].$ Because these integrals have the property $I_m[a^2u(ax)] = a^{2m-1} I_m[u(x)]$ this immediately gives us the equality
$$ I_m[-n(n+1) \mbox{sech}^2 x] = I_m[-2 \mbox{sech}^2 x] \sum_{k=1}^{n} k^{2m-1}.$$
Now to derive the formula (\ref{KdV_Faulhaber}) we need only to show that $I_m[-2 \mbox{sech}^2 x] = (-1)^m 2^{2m}/(2m-1)$, which can be done in various ways. 

All this leads to the formula (\ref{Fsol}), which completes the proof of Theorem 1.
In fact, one can prove a slightly stronger result:
\begin{prop} 
The elliptic Faulhaber polynomial has the form  
$$\mathcal{F}_m(\lambda)= -\frac{4}{2m-1} (g_1^{m-1}\xi - \frac{1}{2}g_1^{m-2}g_2\omega ) F_{m}(\lambda) + \dots $$
where $F_m(\lambda)$ is the usual Faulhaber polynomial and the dots mean the terms of lower 
order in $g_1.$

\end{prop}

There are two more interesting special cases: {\it lemniscatic} when $g_1 = g_3 = 0$ and {\it equianharmonic} when $g_1 = g_2 =0.$ Both of them are specialisations of the reduced version of the elliptic Bernoulli polynomials, which we are going to look at  in more detail in the next section.


\section{The reduced elliptic Faulhaber polynomials.}
 
The reduced form of the elliptic Faulhaber polynomials corresponds to $g_1=0:$  
$$F_m^W(\lambda; g_2, g_3, \omega, \eta)= \mathcal{F}_m(\lambda; 0, g_2, g_3, \omega, \eta).$$
In that case the function $\wp_*$ becomes simply the standard Weierstrass elliptic function $\wp(z; g_2,g_3),$ so by definition
the reduced elliptic Faulhaber polynomials are
\begin{equation}
\label{W}
F_m^W(\lambda | \Gamma, \gamma) = \oint_\gamma T_{m}[2
\lambda \wp(z)] dz.
\end{equation} 
where  $\gamma$ is any cycle on the elliptic curve $\Gamma$ given by the algebraic relation $$Y^2 = 4X^3 - g_2 X - g_3,$$ which does not pass through the poles of the $\wp$-function.

{\bf Remark.} One can consider also another reduction replacing the Weierstrass function $\wp$ by the Jacobi elliptic function $k^2 sn^2,$ appearing in the Jacobi form of the Lam\'e equation \cite{Whittaker}.  The corresponding {\it Jacobi reduced version of the elliptic Bernoulli polynomials}
can be defined by setting $g_3=0:$
$$F_m^J(\lambda; g_1, g_2, \omega, \xi)= \mathcal{F}_m(\lambda; g_1, g_2, 0, \omega, \xi).$$
It corresponds to the elliptic curves $\Gamma$ with a chosen half-period and has the advantage that one can easily see the hyperbolic limit ($g_2=0$). However we prefer the Weierstrass version as it is more canonical and standard in the mathematical literature.


\begin{theorem} 
The elliptic Faulhaber polynomials  $F_m^W$ have the form 
$$F_m^W = A_m (\lambda; g_2, g_3) \omega + B_m (\lambda; g_2, g_3) \eta, $$
$$ A_m = \sum A^{(m)}_{k,l}(\lambda) g_2^k g_3^l, \quad B _m = \sum B^{(m)}_{k,l}(\lambda) g_2^k g_3^l,$$ where the sum is taken over all non negative  integers $k,l$ satisfying $2k+3l=m$ and $2k+3l=m-1$
respectively and $A^{(m)}_{k,l}(\lambda), B^{(m)}_{k,l}(\lambda)$ are some polynomials in $\lambda$ of degree $m$ with rational coefficients and having double zero at $\lambda=0.$
\end{theorem}

The explicit form of the first 8 polynomials can be found in the Appendix. We are going to look in more detail at the highest and the lowest degree coefficients
in $\lambda$ of the polynomials $F_m^W.$ 

The highest coefficient of $F_m^W$ is proportional to $K_n=\oint \wp(z)^n dz.$ Setting  $g_1=0$ in the proposition 2 from the previous section we have 

\begin{prop} The integrals $ K_n $ satisfy the following 2 term recurrence relation of third order
\[(8n-4)K_n = (2n-3)g_2 K_{n-2} + (2n-4)g_3K_{n-3}\]
\label{recurrent rel.}
with the initial data $ K_0 =  2\omega, K_1 = - 2 \eta, K_2 = \frac{1}{6} g_2 \omega.$
They have the form
\begin{equation}
\label{km}
K_m = A^{(m)} (g_2, g_3) \omega - B^{(m)} (g_2, g_3) \eta,
\end{equation}
where $A^{(m)} (g_2, g_3),  B^{(m)} (g_2, g_3)$ are polynomials with positive rational coefficients.
\end{prop}



There is no explicit way to solve this recurrence (except for the lemniscatic and equianharmonic cases, see below), so we would like to discuss here another way to compute $K_n,$ going back to the classical work by Halphen \cite{Halphen}
(see chapter VII page 203).

It is based on the following relation between powers of $\wp$ and its even derivatives:
\begin{equation}
\wp^n= B_n^{(n)} + \sum_{r=0}^{n-1} \frac{B_r^{(n)}}{(2n-2r-1)!} \wp^{(2n-2-2r)},
\label{power of weierstrass}
\end{equation} 
where $B_r^{(n)}$ are coefficients given by the {\it Halphen recurrence relation} (see \cite{Halphen, Tannery}) 
\begin{equation}
 B_r^{(n)} =\frac{(2n-2r-2)(2n-2r-1)}{(2n-2)(2n-1)}B_r^{(n-1)} + \frac{2n-3}{4(2n-1)}B_{r-2}^{(n-2)}g_2 + \frac{n-2}{2(2n-1)}B_{r-3}^{(n-3)}g_3
\label{Halphen relation}
\end{equation}
with $n>0$, $r=0,...,n$ and with  $B_r^{(n)}=0$ for $r<0$ or $r>n$, $B_0^{(n)} =1$ and $B_1^{(n)} =0$ , for any $n.$  By construction $B_r^{(n)}$ are polynomials in $g_2,g_3$ with rational positive coefficients. We will call them {\it Halphen coefficients;}
see the first of them in the Table 1.

 \begin{table}[htdp]
\caption{The Halphen coefficients $B_r^{(n)}$ }
\begin{center}
\begin{tabular}{|c|c|c|c|c|c|c|c|} \hline 
$n$ & 0&1&2&3&4&5&6 \\ \hline \hline  r=0&1&1&1&1&1&1 &1 \\ r=1&0&0&0&0&0&0&0\\ r=2&0&0&$\frac{1}{12}g_2$&$\frac{3}{20}g_2$&$\frac{1}{5}g_2$&$\frac{1}{4}g_2$&$\frac{3}{10}g_2$ \\ r=3&0&0&0&$\frac{1}{10}g_3$&$\frac{1}{7}g_3$&$\frac{5}{28}g_3$&$\frac{3}{14}g_3$ \\
r=4&0&0&0&0&$\frac{5}{336}g_2^2$&$\frac{7}{240}g_2^2$&$\frac{17}{400}g_2^2$\\r=5&0&0&0&0&0&$\frac{1}{30}g_2g_3$&$\frac{87}{1540}g_2g_3$\\r=6&0&0&0&0&0&0&$\frac{15}{4928}g_2^3+\frac{1}{55}g_3^2$\\ \hline
\end{tabular}
\end{center}
\label{Halphen coefficients }
\end{table}

This leads to the following expression for $K_n$ in terms of Halphen coefficients: 
\begin{equation}
K_n =\oint \wp^n (z)dz= 2B_n^{(n)} \omega -2 B_{n-1}^{(n)} \eta. 
\label{KHal}
\end{equation}
Indeed, since the complete integrals of the total derivatives are zero, the only non-zero contribution comes from the constant  and $\wp$ terms in the right-hand side of the relation (\ref{power of weierstrass}).
This is of course related to the general de Rham theorem, which implies that on the elliptic curve $\Gamma$ any differential $\wp^n(z)dz$ is cohomological to a linear combination of $dz$ and $\wp(z)dz.$ 

Note that these particular Halphen coefficients $B_n^{(n)}$ and $B_{n-1}^{(n)}$ satisfy the same recurrence relation (\ref{recurrent rel.}) as $K_n$: $$\alpha_n = \frac{2n-3}{4(2n-1)}g_2 \alpha_{n-2}+ \frac{n-2}{2(2n-1)}g_3\alpha_{n-3}$$  but with  initial conditions $\alpha_{-1}=0, \alpha_0=1, \alpha_1=0 $  for $\alpha_n=B_{n}^{(n)} $ and $\alpha_{-1}=0,\alpha_0=0,\alpha_1=1$ for $\alpha_n=B_{n-1}^{(n)} .$

In the lemniscatic case when $g_3=0$ this recurrence equation can be easily solved and leads to the following formula for $K_n$:
\begin{equation}
K_n= 2\frac{n!}{(2n)!} \prod_{k=1}^{[\frac{n-1}{2}]}(4k-2r+1)^2 g_2^{[\frac{n}{2}]}((1-r)\omega -2r\eta).
\label{Klemn}
\end{equation}

Similarly in the equianharmonic case when $g_2=0,$ we have three cases depending on the congruence of $n$ modulo 3:

\[for ~ n\equiv 0 ~ [3],~~~ K_n= \frac{1}{2^{[\frac{n}{3}]-1}} \prod _{k=1}^{[\frac{n}{3}]} \frac{n-3k+1}{2n-6k+5} g_3^{[\frac{n}{3}]} \omega,\]

\[ for ~ n\equiv 1 ~ [3],~~~ K_n= -\frac{1}{2^{[\frac{n}{3}]-1}} \prod _{k=1}^{[\frac{n}{3}]} \frac{n-3k+1}{2n-6k+5} g_3^{[\frac{n}{3}]} \eta,\]

\[ for ~ n\equiv 2 ~ [3],~~~ K_n= 0.\]

In the next section we analyse the coefficients of the elliptic Faulhaber polynomials at the lowest degree $\lambda^2$, which can be considered as an elliptic generalisation of the Bernoulli numbers.

 \section{The elliptic Bernoulli numbers \label{elliptic Bernoulli}}
 
We define the {\it elliptic Bernoulli numbers} $\mathcal{B}_{2n} = \mathcal{B}_{2n}(g_2, g_3, \omega, \eta)$ as the following integrals
\begin{equation}
\mathcal{B}_{2n} =  \frac{1}{2} \oint _\gamma(\frac{d^{n-1}}{dz^{n-1}} \wp(z; g_2,g_3) )^2 dz. 
\label{defJ}
\end{equation}
They are  simply related to the lowest degree coefficient at $\lambda^2$ in the reduced elliptic Faulhaber polynomial $F_m^W(\lambda),$ which is equal to  $8 \mathcal{B}_{2m-2}.$
Note that the coefficient at lowest degree $\lambda^2$ in $\mathcal{F}_m(\lambda)$ for $m>1$ can be obtained by replacing the coefficients $g_2, g_3, \eta$ in $ \mathcal{B}_{2m-2}$ by $g_2 + \frac{1}{12} g_1 ^2$ and $  g_3 +\frac{1}{12} g_1 g_2 + \frac{1}{216}g_1^3, \xi + \frac{g_1}{12} \omega$ respectively.

The terminology is justified by the fact that on the {\it discriminant,} where two roots of the polynomial $4X^3 - g_2 X - g_3$ collide, these numbers reduce to the classical (even) Bernoulli numbers:
$$B_0 = 1, \, B_2 = \frac{1}{6}, \, B_4 = - \frac{1}{30}, \, B_6 =  \frac{1}{42}, \, B_8 = -  \frac{1}{30}, \, B_{10} =  \frac{5}{66}, \, B_{12}= -\frac{691}{2730},...$$
Recall that all the odd Bernoulli numbers except $B_1 = -1/2$ are zero.

\begin{prop}  For $n>1$ the specialisation of the elliptic Bernoulli numbers on the discriminant is
\begin{equation}
\mathcal{B}_{2n} (\frac{1}{12} g_1^2,\, \frac{1}{216} g_1^3, \,\omega, \, \xi + \frac{1}{12} g_1 \omega) = - B_{2n} \, g_1^n \xi,
\label{discr}
\end{equation}
where $B_{2n}$ are the usual Bernoulli numbers.
\end{prop}

This follows from Theorem 1 and the relation between the coefficients of  $F_m^W$ and $\mathcal{F}_m$ mentioned above. One can show this also more directly using the integral formula
for the Bernoulli numbers found in \cite{GV}.

Let us look now at the properties of the elliptic Bernoulli numbers in the general case.

\begin{prop}  The elliptic Bernoulli numbers  satisfy the following recurrence relation:
\[ \mathcal{B}_{2n} = (-1)^{n-1}(2n-1)! [B_{n+1}^{(n+1)} \omega -  (B_{n}^{(n+1)}  -  B_n^{(n)})\eta] - (2n-1)!\sum_{r=2}^{n-1}\frac{(-1)^{r} B_r^{(n)}}{(2n-2r-1)!} \mathcal{B}_{2n-2r},\]
where $B_r^{(n)}$ are the Halphen coefficients.
\end{prop}

Indeed, integrating by parts, we have  
 $\oint (\wp(z)^{(n-1)})^2 dx= (-1)^{n-1} \oint \wp(z)^{(2n-2)}\wp(z) dz,$ so 
 \begin{equation}
 \label{triv}
 \mathcal{B}_{2n} = \frac{(-1)^{n-1}}{2} \oint \wp(z)^{(2n-2)}\wp(z) dz.
 \end{equation}
  Re-arranging the formula  (\ref{power of weierstrass}) gives
\[ \wp^{(2n-2)}=(2n-1)![\wp^n - \frac{B_2^{(n)}}{(2n-5)!} \wp^{(2n-6)}-...-\frac{ B_{n-2}^{(n)}}{3!} \wp^{(2)} - B_{n-1}^{(n)} \wp - B_n^{(n)}].\] 

Multiplying the above formula by $\wp(z)$ and taking the contour integral over a cycle  gives
\begin{equation}
\mathcal{B}_{2n} = (-1)^{n-1} \frac{(2n-1)!}{2}[\oint \wp^{n+1}(z)dz - \oint B_n^{(n)}\wp(z)dz] - (2n-1)! \sum_{r=2}^{n-1}\frac{(-1)^{r} B_r^{(n)}}{(2n-2r-1)!} \mathcal{B}_{2n-2r}.
\label {elliptic Bernoulli 1}
\end{equation}
Using the relation $K_{n+1} = \oint \wp^{n+1}(z) dz= 2 B_{n+1}^{(n+1)} \omega - 2 B_{n}^{(n+1)} \eta$ and $\oint \wp(z)dz = -2 \eta,$ we obtain the proposition.

The recurrence relation in this proposition can be considered as an elliptic analogue of the recurrence relation for the usual Bernoulli numbers:
\[\sum_{j=0}^m {m+1 \choose j} B_j =0,\]
where ${m+1 \choose j}$ are the binomial coefficients.

There is another way to compute the elliptic Bernoulli numbers if we assume that we know already the 
integrals $K_n.$ Let us use the following very convenient notation:
for a polynomial $D(x) = \sum_{j=0}^N d_j x^j$ and a sequence $K_0, K_1, \dots $ define
$D[K]$ as follows $$D[K] =  \sum_{j=0}^N d_j K_j.$$
Let  $D_n = x A_{n-1} (x; g_2, g_3)= x A^*_{n-1} (x; 0, g_2, g_3),$ where  $A^*_k$  are the polynomials defined in Lemma \ref{p_*^{(n)}} by the recurrence relation (\ref{recur}). Then from (\ref{triv}) we immediately have 
\begin{prop}  The elliptic Bernoulli numbers  can be expressed through $K_n$ as follows:
\[ \mathcal{B}_{2n} = \frac{(-1)^{n-1}}{2} D_n[K].\]
\end{prop}
Here are the first few elliptic Bernoulli numbers:
$$ \mathcal{B}_{2} =   3 g_2 \omega, \quad \mathcal{B}_{4} =  - \frac{3}{5} g_3 \omega + \frac{2}{5} g_2 \eta, \quad   \mathcal{B}_{6} =  \frac{2}{7} g_2^2 \omega - \frac{36}{7} g_3 \eta, $$
$$ \mathcal{B}_{8} =  - \frac{36}{5} g_2g_3 \omega + \frac{24}{5} g_2^2 \eta, \quad  \mathcal{B}_{10} =  \frac{72}{11}( g_2^3 + 18 g_3^2) \omega - \frac{2160}{11} g_2g_3 \eta,$$
$$ \mathcal{B}_{12} =  - \frac{298512}{455}g_2^2 g_3  \omega  + \frac{2592}{455} (49 g_2^3 +750 g_3^2)  \eta, \quad 
\mathcal{B}_{14} =   216 g_2(g_2^3+36g_3^2)  \omega - 9072 g_2^2 g_3  \eta, $$
$$\mathcal{B}_{16} =  - \frac{15552}{85} g_3(1039g_2^3+4500g_3^2)  \omega + \frac{10368}{85}g_2( 539g_2^3+18000g_3^2)   \eta.$$

The special lemniscatic ($g_3 = 0$) and equianharmonic  ($g_2=0$) cases can be seen straight from here, but we can not present a closed formula for $\mathcal{B}_{n}$ for general $n$.

{\bf Remark.} In the classical case the coefficients of the Faulhaber polynomials $F_m(\lambda)$ can be computed from the fact that in the corresponding polynomial $F_m(\frac{x^2+x}{2})= \frac{1}{2m} (B_{2m}(x+1) - B_{2m}) = x^{2m-1} + \frac{1}{2m} (B_{2m}(x) - B_{2m}), $
all the coefficients at odd powers of $x$ except $x^{2m-1}$ are zero (see \cite{K}). 
This is related to the fact that the coefficients of the Bernoulli polynomials are proportional to the Bernoulli numbers and all the odd Bernoulli numbers except $B_1=-1/2$ are zero.
A simple check shows that this is not true in the elliptic case:  already for $\mathcal{F}_3^*$ the coefficient
at $x^3$ in $\mathcal{F}_3^*(\frac{x^2+x}{2})$ is $g_2\xi -2 g_3 \omega$ (which is of course zero if $g_2=g_3=0$ but not in general). This means that the elliptic Bernoulli numbers do not play the same role in the elliptic case as they do in the usual one.

It is worthy to mention that there is another elliptic generalisation of Bernoulli numbers: the so-called {\it Bernoulli-Hurwitz numbers} $BH_{2k}$ \cite{Hurwitz, Katz, Onishi}. They are defined as 
\[ BH_{2k} = 2k(2k-2)! c_k ,\] where $c_k$ are the coefficients of the Laurent series of the Weierstrass $\wp$-function at zero :
\[ \wp(z)=\frac{1}{z^2} + c_2 z^2 + c_3 z^4+...\]
These coefficients $c_k$ satisfy the recurrence relation $$c_k= \frac{3}{(2k+1)(k-3)} \sum_{m=2}^{k-2} c_m c_{k-m}$$ for $k\geq 4$ with $c_2 = g_2/20, c_3 = g_3/28$ (see \cite{Whittaker}).

The Bernoulli-Hurwitz numbers are related to Halphen coefficients in the following way.
Recall that the {\it principal part} of a Laurent series is the sum of the terms with negative powers.
Since the principle part of $\wp^{(2m)}$ is $\frac{(2m+1)!}{z^{2m+2}},$ we have from the relation (\ref{power of weierstrass}) the following
\begin{prop}  The principal part of the Laurent series $\wp^n = (\frac{1}{z^2} + c_2 z^2 + c_3 z^4+.....)^n$ is $$\sum_{r=0}^{n-1} \frac{B_r^{(n)}} {z^{2n-2r}}.$$ 
\end{prop}
Since $(\frac{1}{z^2} + c_2 z^2 + c_3 z^4+.....)^n= \frac{1}{z^{2n}} + \frac{0}{z^{2n-2}} + \frac{nc_2}{z^{2n-4}}+ \frac{nc_3}{z^{2n-6}}+\frac{nc_4 +\frac{n(n-1)}{2}c_2^2}{z^{2n-8}}+\frac{nc_5 +n(n-1)c_2c_3}{z^{2n-10}}..... ,$
this allows us to express the Bernoulli-Hurwitz numbers through Halphen coefficients and vice versa:
$B_0^{(n)} = 1, B_1^{(n)}= 0,$ 
\[B_2^{(n)} = n c_2 = n g_2/20= \frac{n}{8}BH_4,\]
 \[B_3^{(n)} = n c_3 = n g_3/28=\frac{n}{144}BH_6,\] 
 \[B_4^{(n)} = nc_4 +\frac{n(n-1)}{2}c_2^2= \frac{n(3n-1)}{2} c_4= \frac{n(3n-1)}{11520} BH_8,\]
  \[B_5^{(n)}=nc_5 +n(n-1)c_2c_3=\frac{n}{3}(22n-19)c_5 = \frac{n(22n-19)}{1209600} BH_{10}.\]
  Note that these expressions for $B_r^{(n)}$ are only valid when $r<n.$

\section{Application:  the density of states of the Lam\'{e} operators}

The  Lam\'{e} operator $L_n= - \frac{d^2}{dz^2} + n(n+1) \wp(z)$ has been a classical object of investigation since the XIX-th century although its spectral properties were appreciated much later. In 1940  Ince \cite{Ince} proved that the Lam\'{e} operator (on the real line shifted by an imaginary half-period) for integer $n$ has  exactly $n$ gaps in its spectrum. The ends of the gaps are given by the zeroes of certain polynomials $R_{2n+1}(E)$, which we will call {\it spectral.} The problem of finding these polynomials was investigated in the classical works by Hermite and Halphen (see \cite{Whittaker}) but still continues to be of interest (see e.g. recent papers by Belokolos and Enolski \cite{BE} and Takemura \cite{Take1, Take2} and references therein).

In this section we will discuss the density of states for the Lam\'{e} operators, which is one of the most important notions in the spectral theory of the Schr\"odinger operator with periodic potential  (see e.g.
\cite{Shenk}). In the finite-gap case, in particular for the Lam\'e operator $L_n,$ the density of states
has the form  
\begin{equation}
\label{den}
\rho(E)=\frac{1}{2 \pi }  \frac{P_n(E)}{ \sqrt{R_{2n+1}(E)}},
\end{equation} 
where $ P_n(E) = E^n + a_1 E ^{n-1} + a_2 E^{n-2} +...+ a_n $ for some coefficients $a_1, a_2,....,a_n$ and $R_{2n+1}(E) = \prod_{i=0}^{i=2n} (E- E_i)=\sum_{k=0}^{k=2n+1} b_kE^{2n+1-k} $ is the spectral polynomial. We are going to explain how to find the coefficients $a_k$ as a function of $n$ in terms of the elliptic Faulhaber polynomials and the coefficients of the spectral polynomials, which were investigated in \cite{GV1}. In particular we prove the Theorem 2 from the Introduction.



Let us start with the known results about the density of states for the Schr\"odinger operator $L= - \frac{d^2}{dx^2} + u(x) $ on $\mathbb{R}$ with periodic potential $u(x)$ of period $T$ (see e.g. \cite{Shenk, Nov}). The spectrum of the operator $L$ is known to be continuous and has a band structure with in general an infinite number of gaps. Consider the restriction $L^{(R)}$ of $L$ on $(-R,R)$, for $R>0$, with Dirichlet boundary conditions at $\pm R.$ The spectrum of $L^{(R)}$ is then discrete. Denote the eigenvalues as $ \mathcal{E}_n(L^{(R)})$ and define the {\it integrated density of states} $N(E)$ as the limit
\[ N(E)= \lim_{R \rightarrow \infty} \frac{\sharp \{ \mathcal{E}_n(L^{(R)} < E \}}{2R} ,\]
where $\sharp \{  \mathcal{E}_n(L^{(R)}\} $ is the counting function of the discrete spectrum of $ L^{(R)}.$
It is known that such limit exists and has the following asymptotic expansion as $E \rightarrow \infty:$
\begin{equation}
 N(E) = \frac{ \sqrt E}{\pi} -\frac{1}{2 \pi T \sqrt E}  \sum_{k=0}^{\infty}   \frac{I_{k+1}[u]}{(4E)^{k}},  
\label{integrated density}
\end{equation}
where $I_k[u] = \int_T T_{k}(u, u_, u_2, \dots, u_{k-2}) dx$ are the KdV integrals (\ref{Kdvint}) evaluated over period $T$.

The {\it density of states} is defined as the derivative of the integrated density of states:
\[ \rho(E) = \frac{dN(E)}{dE}.\]
It is known (see \cite{Shenk}) that for the periodic potentials $u(x)$ the integrated density of states is related in a very simple way to the so-called {\it quasi-momentum} $p(E)$:
\begin{equation}
 N(E) = \frac{p(E)}{\pi}
\label{relation density momentum}
\end{equation}
Recall that the quasi-momentum $p(E)$ appears naturally in relation with the Bloch (Floquet) solutions of the Schr\"odinger equation
$L \psi = E \psi,$ which have the form $ \psi_{E}(x) = \exp^ {\imath p(E)x} \phi(x),$ $\phi(x)$ being a periodic function of period $T.$

If $u(x)$ is a finite gap potential with $n$ gaps then the derivative of the quasi-momentum $p(E)$  has the form \cite{Nov}:
\[ \frac{dp(E)}{dE} = \frac{1}{2  } \frac{P_n(E)}{ \sqrt{R_{2n+1}(E)} } \]
where $ P_n(E) = E^n + a_1 E ^{n-1} + a_2 E^{n-2} +...+ a_n $ for some coefficients $a_1, a_2,....,a_n$ and $R_{2n+1}(E) = \prod_{i=0}^{2n} (E- E_i) $ where $E_0, E_1,...,E_{2n}$ are the end points of the spectrum intervals of $L.$

Thus in that case, the density of states has the following form
\begin{equation}
\rho(E) =\frac{1}{2 \pi }  \frac{P_n(E)}{ \sqrt{R_{2n+1}(E)}}.  
\label{density of states}
\end{equation}
Note that from (\ref{integrated density}) it follows that it has the following expansion at infinity
\begin{equation}
\label{expan}
\rho(E) = \frac{1}{ 2\pi \sqrt E} +\frac{1}{ \pi T \sqrt E}  \sum_{k=1}^{\infty}  (2k-1)\frac{I_{k}[u]}{(4E)^{k}}. 
\end{equation}

Consider the case of the Lam\'{e} operator $L_n= - \frac{d^2}{dx^2} + n(n+1) \wp.$ 
The integrals $I_{k}[u]$ are then the  elliptic Faulhaber polynomials $ F_{k}^W(\lambda)$, where $\lambda = \frac{n(n+1)}{2}$ and $ \quad T = 2 \omega.$ 
Thus, the  high-energy asymptotics for the densities of states of the Lam\'{e} operator can be given in terms of the elliptic Faulhaber polynomials as follows: 
\begin{equation}
N_n(E) =  \frac{ \sqrt E}{\pi} -\frac{1}{4 \pi \omega \sqrt E}  \sum_{k=0}^{\infty} \frac{F_{k+1}^W(\lambda)}{(4E)^{k}} , \quad  \lambda = \frac{n(n+1)}{2}
\label{Lame integrated density }
\end{equation}
\begin{equation}
\rho_n(E) = \frac{1}{2 \pi \sqrt E} [1+ \frac{1}{2\omega} \sum_{k=1}^{\infty} \frac{ 2k-1}{2^{2k-1}}\frac{ F_{k}^W(\lambda)}{ E^{k}}] , \quad  \lambda = \frac{n(n+1)}{2}.
\label{Lame density of states}
\end{equation}

Thus in principle, if we know all the Faulhaber polynomials then we know the density of states of Lam\'e
operators $L_n$ {\it for all $n$}. In particular, one can use a general procedure based on the theory of continued fractions going back to Stieltjes to reconstruct it. However it does not look effective enough,
so we will take a different approach.

First, we express the coefficients $b_k$ of the spectral polynomials using the so-called {\it elliptic Bernoulli polynomials}  \cite{GV1}.  More precisely, in \cite{GV1} we have proved that the coefficients $b_k$ of the Lam\'e spectral polynomial $$R_{2n+1}(E) = \prod_{i=0}^{2n} (E- E_i) = E^{2n+1} + b_1 E^{2n}+ b_2 E^{2n-1}+ ...+ b_{2n+1} $$ are  polynomials in $n, g_2, g_3$ with rational coefficients. The degree of $b_k$ in $n$ is $[\frac{5k}{2}]:$
\[ b_1 =0\]
\[b_2= -\frac{g_2}{120}n(n+1)(2n-1)(2n+1)(2n+3)\]
\[b_3= - \frac{g_3}{840}n(n+1)(2n-3)(2n-1)(2n+1)(2n+3)(2n+5)\]
\[ b_4 =  \frac{g_2^2}{201600}n(n-1)(n+1)(2n-1)(2n+1)(2n+3)(56n^4 +76 n^3 -94n^2+201n+630).\]  

Re-writing the high energy asymptotics of the density of states of the Lam\'e operator $L_n$  in the form
$$\frac{1}{2 \pi }  \frac{P_n(E)}{ \sqrt{R_{2n+1}(E)}}
= \frac{1}{2 \pi \sqrt E} [ 1 + \frac{a_1}{E} + ( a_2 -  \frac{  b_2}{2})\frac{1}{E^2} + (a_3 - \frac{a_1b_2}{2} - \frac{b_3}{2})\frac{1} {E^3}+ ...$$
and equating the coefficients term by term with the expansion (~\ref{Lame density of states}) gives
\[a_1= \frac{ F_1^W(\lambda)}{4w} = - \frac{\eta}{\omega} \lambda =-  \frac{\eta}{\omega} \frac{n (n+1)}{2}, \]
\[ a_2 =   \frac{ b_2}{2} + \frac{3}{16} \frac{ F_2^W(\lambda)}{\omega}, \]
\[a_3 = \frac{a_1b_2}{2} +\frac{b_3}{2}+ \frac{5}{2^6} \frac{ F_3^W(\lambda) } {\omega},\]
\[a_4 =\frac{a_2 b_2}{2} -  \frac{3b_2 ^2}{8} + \frac{a_1 b_3}{2} +\frac{b_4}{2} +\frac{7}{2^8} \frac{F_4^W[\lambda] } {\omega}.\]
From this we find recursively
\[a_1=\frac{n (n+1)}{2} \bar{\wp},\]
\[ a_2 = -\frac{g_2}{480}(n-1) n(n+1)(6+25 n +16 n^2),\]
\[a_3= -\frac{g_3}{3360} (n-2)(n-1)n(n+1)(45 + 243n + 247 n^2+ 64 n^3) - \frac{g_2\bar{\wp}}{960}(n-2)(n-1)n^2(n+1)^2(27 + 16 n),\]
\[a_4 =  \frac{g_2^2}{3225600}(n-3)(n-2)(n-1)n(n+1)(-2520-12942n-10315n^2+4565n^3+6880n^4+1792n^5)\] \[ -\frac{g_3\bar{\wp}}{13440}(n-3)(n-2)(n-1)n^2(n+1)^2(600+563n+128n^2),\]
where $ \bar{\wp} =-\frac{\eta}{\omega}$ is the average of $\wp(x)$ over a period. 
Note that the coefficient  $a_k$ is divisible by $(n+1)n(n-1)(n-2)...(n-k+1),$ which is related to the fact that
$L_n$ has $n$ gaps for integer $n$. All this implies Theorem 2.

{\bf Remark.} The differential $\rho(E) dE =\frac{1}{2 \pi }  \frac{P_n(E) dE}{ \sqrt{R_{2n+1}(E)}}$
has a simple algebro-geometric meaning: it is an Abelian differential with a second order pole at infinity normalised by the condition that all the periods over gaps ($b$-periods) are zero (see e.g. Novikov et al \cite{Nov}). This means that once the equation of the spectral curve is known one can find the coefficients of the numerator by solving a linear system of equations with the coefficients being some standard hyperelliptic integrals on this curve. From our results it follows that these coefficients actually can be expressed polynomially in terms of the standard {\it elliptic} integrals and the parameters of the initial elliptic curve. This is of course related to the {\it reduction problem} for the Lam\'e spectral curves, see e.g. \cite{BE, Take1}.

One can see the explicit form of the first 7 coefficients and the first 5 Lam\'e densities of states in the Appendix. They are in a good agreement with Belokolos-Enolski and Takemura calculations \cite{BE, Take1}, who used a modified version of the classical approach to this problem.

 \section{Some open questions and conjectures.}
 
The coefficients of the classical Faulhaber polynomials are known to have  alternating signs
(see \cite{K}). We conjecture that the same is true for the reduced elliptic Faulhaber polynomials.
More precisely, recall that in the Weierstrass form these polynomials have the form
 $F_m^W = A_m (\lambda; g_2, g_3) w + B_m (\lambda; g_2, g_3) \eta, $ where
$$ A_m = \sum A^{(m)}_{k,l}(\lambda) g_2^k g_3^l, \quad B _m = \sum B^{(m)}_{k,l}(\lambda) g_2^k g_3^l.$$ 
 
\begin{conj} The coefficients of the polynomials  $A^{(m)}_{k,l}(\lambda)$ and  $B^{(m)}_{k,l}(\lambda)$ have alternating signs. 
\end{conj}

For the first 8 polynomials one can check this from the explicit form given in the Appendix. We have checked that a similar property holds for the Jacobi reduced versions (but not for the general case, see e.g. $\mathcal{F}_4$ in the Introduction). 
An ideal proof would be to find a combinatorial interpretation for these coefficients (cf. Gessel and Viennot \cite{Gessel} and Knuth \cite{K} in the classical case). It might be easier to prove it first for the elliptic Bernoulli numbers (c.f. Proposition 4 above).

\begin{conj} The elliptic Bernoulli numbers $\mathcal{B}_{2m}$ have the form
$$\mathcal{B}_{2m} = (-1)^{m-1} (\hat A^{(m)}(g_2, g_3) w - \hat B^{(m)}(g_2, g_3) \eta),$$
where the polynomials $\hat A^{(m)}(g_2, g_3),  \hat B^{(m)}(g_2, g_3)$ have {\bf positive} rational coefficients.
\end{conj}

Another interesting question is the behaviour of the elliptic Faulhaber polynomials on a real line.
Actually it is more instructive to look at the corresponding polynomials $\mathcal{F}_m(\frac{x^2+x}{2})$
as functions of $x$ rather than $\lambda=\frac{x^2+x}{2}.$ Indeed, we know that the integer values $x=n$ should play a special role here, being related to the "finite-gap" values of the parameter in the Lam\'e equation. Recall also that in the usual case these polynomials coincide with $\frac{1}{2m} (B_{2m}(x+1) - B_{2m}),$
so their graphs are simply shifted graphs of the corresponding Bernoulli polynomials.

\begin{conj} As $m$ tends to $\infty$ for $x$ in a finite interval on the real line
$$\frac{\mathcal{F}_m(\frac{x^2+x}{2})}{2 \mathcal{B}_{2m-2}} \rightarrow \frac{1-\cos 2\pi x}{2 \pi^2}.$$
\end{conj}

In the hyperbolic limit $g_2=g_3=0$ (i.e. for the usual Bernoulli polynomials) it is known to be true (see e.g. \cite{VW}).
The normalisation constant $2 \mathcal{B}_{2m-2}$ is chosen to guarantee the correct second derivative (chosen to be 1) at zero. Another justification for this conjecture is that the asymptotic expansion (~\ref{Lame density of states}) should be convergent for integer $n$ (due to the finite-gapness property), which means that the numbers $\mathcal{F}_m(\frac{n^2+n}{2})$ should be relatively very small.

In the Fig. 1 we show the graphs of the normalised polynomials $$\Phi_m(x)=\frac{\mathcal{F}_m(\frac{x^2+x}{2})}{2 \mathcal{B}_{2m-2}}$$ for $m=8$ in the usual and lemniscatic elliptic case $g_1=g_3=0.$ 


\begin{figure}
\centerline{ \includegraphics[width=6cm]{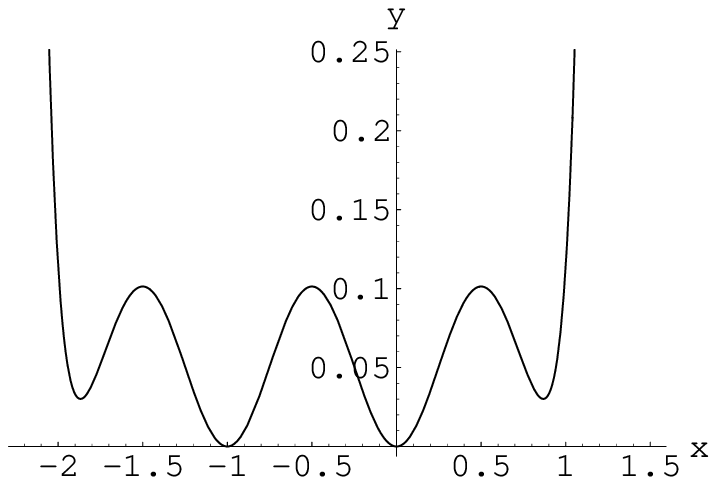} \hspace{10pt}
\includegraphics[width=6cm]{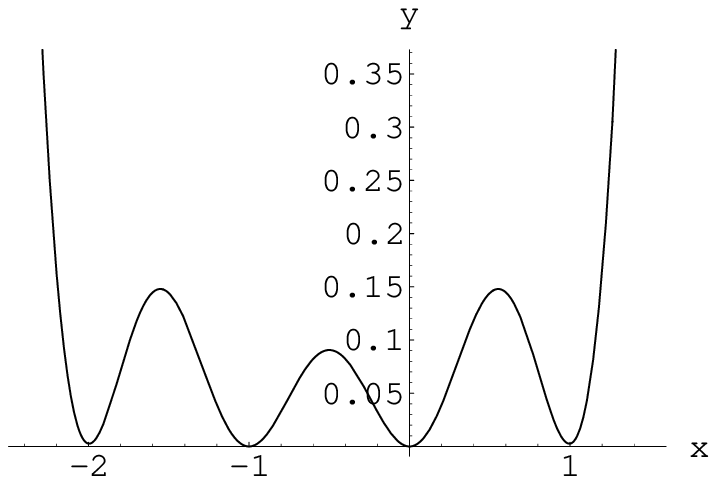} }
\caption{Left: $\Phi_8(x)$ for the usual normalised Faulhaber polynomial, Right: $\Phi_8^{W}(x)$ in the elliptic lemniscatic case} \label{boxflower}
\end{figure}

The elliptic Faulhaber polynomials have further generalisation related to the so-called {\it Treibich-Verdier} potentials \cite{TV}
$$u = m_0(m_0+1)\wp(x) + \sum_{i=1}^{3}m_i(m_i+1)\wp(x-\omega_i),$$
where $\omega_i$ are the half-periods of the Weierstrass elliptic $\wp$-function.
The corresponding KdV integrals $I_k[u]$  are polynomials in  four variables $\lambda_i =  m_i(m_i+1), \, i=0,1,2,3$ with the coefficients depending on $e_1, e_2, e_3, \omega, \eta.$ We are planning to discuss them elsewhere.

We believe also that the elliptic Faulhaber polynomials and elliptic Bernoulli numbers  
are interesting from a number-theoretic point of view and would like to mention the papers
\cite{Katz, Katz2, Onishi} in this relation.
 
 \section{Acknowledgements.}

We are very grateful to Jean-Benoit Bost and Victor Enolski for useful and stimulating discussions.

 \section{Appendices}


 

\subsection{Appendix A}  
{\bf  The first 8 reduced elliptic Faulhaber polynomials $F_k^W$} 

\medskip

The lemniscatic and equianharmonic cases can be read off by putting $g_3=0$ and $g_2=0$ respectively.

\bigskip

\noindent
$F_1^W  =$ \hspace {0.3in}
$\begin {array}{lll}
\vspace{0.1 in}-&   \eta &[4 \lambda ]
\end{array}$ 
\newline \newline
$ F_2^W  =$ \hspace {0.3in}
$\begin {array}{lll}
\vspace{0.1 in}+& g_2 \omega &[\frac{2}{3} \lambda ^2 ]
\end{array}$
  \newline \newline \newline 
$ F_3^W =$ \hspace {0.3in}
$\begin {array}{lll}
\vspace{0.1 in}-& g_2 \eta&[\frac{8}{5} \lambda ^2 ( 3 \lambda -2)] \\
\vspace{0.1 in}+&g_3  \omega  & [\frac{8}{5} \lambda ^2 (2\lambda -3)]    \end{array}$
  \newline \newline \newline 
  $F_4^W =$ \hspace {0.3in}
$\begin {array}{lll}
 \vspace{0.1 in}+& g_2^2  \omega &[\frac{2}{21} \lambda ^2 ( 25 \lambda ^2 - 40 \lambda + 24)] \\ \vspace{0.1 in}-&g_3 \eta  & [\frac{32}{7} \lambda ^2 (5 \lambda ^2 -15\lambda +9)] 
  \end{array}$
  \newline \newline \newline 
  $F_5 ^W  =$ \hspace {0.3in}
$\begin {array}{lll}
\vspace{0.1 in}-& g_2^ 2 \eta&[\frac{8}{15} \lambda ^2 (49 \lambda ^ 3 -140\lambda ^2 +168 \lambda -72)] \\ \vspace{0.1 in} +&g_2 g_3  \omega  & [\frac{16}{15} \lambda ^2 (28 \lambda ^3 - 105 \lambda ^2 + 126\lambda -54)] \
  \end{array}$
  \newline \newline \newline 
    $ F_6^W  =$ \hspace {0.3in}
$\begin {array}{lll}
\vspace{0.1 in}+& g_2^3  \omega &[\frac{4}{11} \lambda ^2 (45 \lambda ^4 - 200 \lambda ^ 3 + 416\lambda ^2 - 400 \lambda +144)]  \\ \vspace{0.1 in}-& g_2 g_3 \eta&[\frac{192}{55} \lambda ^2 ( 87 \lambda ^4 -515 \lambda ^ 3 +1179\lambda ^2 -1206 \lambda +450)] \\  \vspace{0.1 in} +&g_3 ^2  \omega  & [\frac{96}{55} \lambda ^2 (  56 \lambda ^4 - 420 \lambda ^3 + 1197 \lambda ^2 - 1368 \lambda +540)]
 \end{array}$
  \newline \newline \newline 
  $ F_7^W  =$ \hspace {0.3in}
 $\begin {array}{lll}
 \vspace{0.1 in}-& g_2^3 \eta &[\frac{16}{65} \lambda ^2 (847 \lambda ^5 - 5390 \lambda ^4 + 17248 \lambda ^ 3 - 30536\lambda ^2 + 26928 \lambda - 9072)] \\ \vspace{0.1 in}+& g_2^2 g_3  \omega &[\frac{16}{455} \lambda ^2 ( 9526\lambda ^5  -71995 \lambda ^4 + 250404 \lambda ^ 3 - 472428\lambda ^2 + 433224 \lambda - 149256)]  \\ \vspace{0.1 in}-&g_3 ^2 \eta  & [\frac{384}{91} \lambda ^2 (  220 \lambda ^ 5 - 2310 \lambda ^4 +10395 \lambda ^3 - 22770 \lambda ^2 + 22572 \lambda - 8100)]
 \end{array}$
  \newline \newline \newline 
 $F_8 ^W =$ \hspace {0.3in}
  \small
$\begin {array}{lll}
 \vspace{0.1 in}+& g_2^4  \omega &[\frac{2}{21} \lambda ^2 ( 1521\lambda ^6 -13104 \lambda ^5 + 59696 \lambda ^4 -165568 \lambda ^ 3 + 269568 \lambda ^2 - 224640 \lambda + 72576)] \\ \vspace{0.1 in}-& g_2^2 g_3 \eta &[\frac{64}{35} \lambda ^2 ( 2171 \lambda ^6  -22477\lambda ^5  +113295 \lambda ^4 -336492 \lambda ^ 3  + 570492\lambda ^2 - 485784 \lambda + 158760)]  \\ \vspace{0.1 in}+&g_2 g_3 ^2  \omega  & [\frac{64}{5} \lambda ^2 (182 \lambda^6 - 2184 \lambda ^ 5 +12285 \lambda ^4 -38844 \lambda ^3 +68094 \lambda ^2 -58968 \lambda +19440)]
 \end{array}$

\normalsize

 \subsection{Appendix B} 
{\bf The list of the first 7 coefficients $a_k(n)$ of the numerator $P_n(E)$ of the density of the Lam\'e states}  

\medskip

Here we use the notations $\bar{\wp}=-\frac{\eta}{\omega}, \quad U_k(n) = (n+1)n(n-1) \dots (n-k+1), \quad \hat a_k (n) = a_k (n) / U_k(n).$

\medskip

\noindent
$\hat a_1= \frac{1}{2} \bar{\wp},$ \newline \newline
$ \hat a_2 = -\frac{g_2}{480}(6+25 n +16 n^2),$ \newline \newline
$\hat a_3 = -\frac{g_3}{3360} (45 + 243n + 247 n^2+ 64 n^3) - \frac{g_2\bar{\wp}}{960}n(n+1)(27 + 16 n),$ \newline \newline
$\hat a_4 =  \frac{g_2^2}{3225600}(-2520-12942n-10315n^2+4565n^3+6880n^4+1792n^5)-\frac{g_3\bar{\wp}}{13440}n(n+1)(600+563n+128n^2),$
\newline \newline
$\hat a_5  = \frac{g_2g_3}{17740800} (-28350-145305n-98919n^2+130400n^3+185250n^4+78480n^5+ 11264 n^6) +\frac{g_2^2\bar{\wp}}{6451200}n(n+1)(-22050-19707n+3217n^2 +7328 n^3 + 1792 n^4),$ \newline \newline
$\hat a_6 = \frac{g_3^2}{3228825600}  (585728 n^7+6077568n^6+24055710n^5+42381080 n^4+22989372 n^3-21506058 n^2-26135595n-4677750)-\frac{g_2^3}{664215552000}  (4100096 n^8+23905024 n^7+14017296 n^6-192219241 n^5-520562096 n^4-391295859 n^3+180468864 n^2+310981356 n+62868960) + \frac{g_2 g_3 \bar{\wp}}{70963200}n(n+1)(22528n^5+171950n^4+427045n^3+215715n^2-568458n-612360),$ \newline \newline
$\hat a_7 = - \frac{g_2 ^2 g_3}{1549836288000}( 16400384 n^9 +160552704 n^8+520180864 n^7+153752039 n^6-2909673459n^5-6672918014n^4-4259587899n^3+2522284551n^2+3574728990n+681080400) + \frac{g_3^2 \bar{\wp}}{6457651200}n(n+1)
(585728n^6+6709056n^5+29129901n^4+54097586n^3+19647288n^2-62676225n-60031125)- \frac{g_2 ^3 \bar{\wp}}{1328431104000}n(n+1)(4100096n^7+25442560n^6+8937936n^5-279814135n^4-748894946n^3-399303225n^2+770746914n+806818320).$

\medskip

\subsection{Appendix C} 
{\bf The explicit form of the first 5 Lam\'e densities of states}  

\medskip

\begin{itemize}
\item For $n=1$ the potential is $u(x) = 2 \wp(x).$ 

The density of states is well-known in this case:
\[\rho_{Lame}^{n=1} (E)=\frac{1}{2 \pi}\frac{E +\bar{\wp}}{\sqrt{E^3 -\frac{g_2}{4}  E+\frac{g_3}{4}}} =\frac{1}{2 \pi}\frac{E +\bar{\wp}}{\sqrt{(E+e_1)(E+e_2)(E+e_3)}}\]
\item For $n=2$ the potential is $u(x) = 6 \wp(x),$ 
the density of states is
\[\rho_{Lame}^{n=2} (E)= \frac{E^2 +3\bar{\wp} E -\frac{3}{2}g_2}{ \pi \sqrt{4E^5 -21g_2 E^3 -27 g_3 E^2 + 27 g_2^2 E + 81 g_2 g_3}}\]
\item For $n=3$ the potential is $u(x) = 12 \wp(x)$
and the density of states is
\[ \frac{E^3 +6\bar{\wp} E^2 -\frac{45}{4}g_2E  -\frac{135}{4}g_3-\frac{45}{2}g_2\bar{\wp}  }{2 \pi \sqrt {E^7 - \frac{63}{2} g_2 E^5 - \frac{297}{2} g_3 E^4 + \frac{4185}{16} g_2 ^2 E^3 + \frac{18225}{ 8} g_2 g_3 E^2 - \frac{3375}{16} ( g_2 ^3 - 27 g_3 ^2)E}}\]
 \item For $n=4$ the potential is $u(x) = 20 \wp(x)$.
The density of states is
\[\rho_{Lame}^{n=4} (E)= \frac{E^4 +10\bar{\wp} E^3 -\frac{181}{4}g_2E^2  -(\frac{1295}{4}g_3+\frac{455}{2} g_2\bar{\wp})E + \frac{273}{2} g_2^2 - 875 g_3 \bar{\wp}  }{2 \pi \sqrt{(E^3 - 52 g_2 E - 560 g_3)\prod_{k=1}^{k=3}(E^2 - 10 e_k E - 35 e_k^2- 7 g_2)}}.\] 
\item  For $n=5$, the potential is $u(x) = 30 \wp(x)$ and the density of states is
\small
\[ \frac{  E^5 + 15\bar{\wp}E^4 -\frac{531}{4}g_2 E^3 - (\frac{6615}{4}g_3+\frac{4815}{4}g_2\bar{\wp} )E^2 + ( \frac{18117}{8} g_2^2 -\frac{42525}{4}g_3\bar{\wp})E + \frac{178605}{8}g_2g_3 + \frac{13365}{2}g_2^2 \bar{\wp} }{2 \pi \sqrt{(E^2 - 27 g_2 ^2)\prod_{k=1}^{k=3}(E^3 + 15 e_k E^2 +( 315 e_k^2 -132 g_2) E - 675 e_k^3 -540 g_3)}}.\] 
\normalsize
\end{itemize}

\end{document}